\documentstyle[prl,aps]{revtex}
\begin{document}
\title{Solidity of viscous liquids II: Anisotropic flow events}
\draft
\twocolumn
\author{Jeppe C. Dyre}
\address{Department of Mathematics and Physics (IMFUFA), 
Roskilde University, Postbox 260, DK-4000 Roskilde, Denmark}
\date{\today}
\maketitle{}

\begin{abstract}
Recent findings on the displacements in the surroundings of
isotropic flow events in viscous liquids [Phys. Rev. E, in press]
are generalized to the anisotropic case.
Also, it is shown that a flow event is characterized by a 
dimensionless number reflecting the degree of anisotropy.
\end{abstract}


In a previous paper henceforth referred to as (I) \cite{dyr98b},
it was argued that viscous liquids close to the glass transition 
\cite{kau48,har76,bra85,dyr87,ang88,ang91,moh95,nem95,ang96,%
ang97} - where viscosity is roughly $10^{15}$ times larger than
that of, e.g., room temperature water - are more like solids than
like the less viscous liquids studied in standard liquid theory
\cite{han86,boo80}.
The idea \cite{dyr87} that viscous liquids are qualitatively
different from less-viscous liquids is, of course, not new.
It is a rather obvious idea, given the following fact. 
While ``ordinary'' less-viscous liquids have relaxation times in
the picosecond range, i.e., comparable to typical phonon times,
viscous liquids have much longer average relaxation times
(roughly given by Maxwell's expression 
$\tau = \eta / G_{\infty}$, where $\eta$ is the viscosity and
$G_{\infty}$ the instantaneous shear modulus).
This decoupling of relaxation times from phonon times is also
reflected in a decoupling of diffusion constants \cite{maj87}:
For less-viscous liquids the molecular diffusion constant $D$ is
of the same order of magnitude as the transverse momentum
diffusion constant, the dynamic viscosity of Navier-Stokes
equation $\nu\propto\eta$.
However, with increasing viscosity $D$ decreases (roughly as
$\eta^{-1}$ from a simple Stokes-Einstein type argument) while
$\nu$ increases.
At the glass transition $\nu$ is about ${\rm 10^{30}}$ times
larger than $D$.

The average relaxation time increases dramatically upon cooling.
Goldstein has argued \cite{gol69} that already when $\tau$
becomes longer than about 1 nanosecond is there a gradual onset
of typical viscous-liquid behavior.
As noted first by Angell \cite{ang88}, this is roughly at the
temperature below which ideal mode-coupling theory \cite{got92}
breaks down.
It is generally believed that in viscous liquids ``real''
molecular motion beyond pure vibration takes place on the time
scale defined by $\tau$, although inhomogeneities are likely to
give rise to faster relaxations in some parts of the liquid
\cite{per96,moy93,wan96,boh96,cic95,hur95}.
``Real'' motion is rare because it involves overcoming 
energy barriers large compared to $k_BT$ \cite{kau48,gol69}.
The transition itself is a jump between two potential energy
minima, a process that lasts just a few picoseconds.
One thus arrives at the following picture:
Most molecular motion in a viscous liquid is purely vibrational,
real motion is rare and takes place via sudden molecular
rearrangements.
It is interesting to note that this old picture
\cite{kau48,eyr36} has never really been challenged (while the
nature of the energy barrier to be overcome in the transition is
still being debated \cite{dyr98a}).
In fact, extensive computer simulations have now definitively
confirmed the picture \cite{sch99}.

The sudden molecular rearrangements 
\cite{kau48,ada65,gol72,don82,sti88,sch90,cha93}
are referred to as ``flow events'' below.
It is generally believed that flow events are localized in the
sense that only a limited number of molecules experience large
displacements, while all other molecules are only slightly
displaced; the large-displacement molecules involved in a flow
event thus define a ``region'' of the liquid.
Because flow events are rare and molecules most of the time just
vibrate, a viscous liquid looks much like a solid. 
In (I) the small displacements in the surroundings of a flow
event were calculated from solid elasticity theory assuming
spherical symmetry. 
It was shown that the displacement $u$ in the surroundings of a
region is given by (where $r$ is the distances to the region)

\begin{equation}\label{1}
u\propto\frac{1}{r^2}\,.
\end{equation}
The displacement is purely radial.
However, spherical symmetry of flow events is not realistic; when
molecules move from one potential energy minimum to another there
must be some violation of spherical symmetry, even if the
molecules were spheres with only radially dependent interactions.
One is thus lead to ask to whether Eq.\ (\ref{1}) and its
consequences remain valid in the anisotropic case.

As in (I) the starting point is the solidity of viscous liquids
as reflected in the slow ``real'' motion of the molecules.
This fact implies that the average force on any molecule is
extremely close to zero.
In a continuum description, the average force per unit volume is
the divergence of the stress tensor $\sigma_{ij}$, where
$i,j=1,2,3$ are spatial indices.
The condition of average zero force - elastic equilibrium - is
(where $\partial_i = \partial/\partial x_i$ and one sums over
repeated indices)

\begin{equation}\label{2}
\partial_i \sigma_{ij}\ =\ 0\,.
\end{equation}

Linear elasticity theory \cite{l+l} may be applied to the region
surroundings, because the molecular displacements in these
surroundings are small and because there is elastic equilibrium
in the liquid before as well as after a flow event.
Most likely, there are large ``frozen-in'' stresses in the
liquid, but the {\it changes} in the stress tensor induced by one
flow event are small, except in the region itself.
Now, define a sphere centered at the region, large enough that
outside the sphere the flow event induced displacements and
stress tensor changes are so small that linear elasticity theory
applies for the changes.
Imagine all molecules within the sphere being removed and
the forces from these molecules acting on the molecules outside
the sphere replaced by external forces applied to the surface of
the sphere.
This is done before as well as after the flow event. 
The flow event induced displacements of the surroundings can then
be calculated from the change of these external forces.
To do this we first consider the distance dependence of
displacements in an elastic solid when an external force is
applied to just one point.
There is then a continuous flow of momentum into the solid at
that point.
The stress tensor is the momentum current and the mechanical
equilibrium condition Eq. (\ref{2}) is the zero-divergence
equation reflecting momentum conservation.
By considering Gauss surfaces at various distances from the
point, one concludes from Eq.\ (\ref{2}) that the stress tensor
decays as $r^{-2}$, where $r$ is the distance to the point.
Since the stress tensor is formed from first order spatial
derivatives of the displacement $u$, we conclude that $u\propto
r^{-1}$ \cite{note}.
This result applies also when several external forces are applied
to the solid, as long as these forces do not sum to zero.
In our case, the external forces replacing the forces from the
molecules within the sphere do sum to zero:
The forces from the molecules {\it outside} the sphere on those
{\it inside} must sum to zero - otherwise the latter molecules
would start to move.
By Newton's third law, the sum of the forces acting from the
molecules {\it inside} the sphere on those {\it outside} - the
forces that are replaced by external forces - must therefore also
sum to zero.
When the external forces sum to zero, the stress tensor does not
decay as $r^{-2}$ but as $r^{-3}$ (the mathematics behind this
fact is the same as that implying that the electric field from a
charge distribution with zero total charge decays as $r^{-3}$ and
not as $r^{-2}$).
Consequently, since the stress tensor is given as first order
derivatives of the displacement vector we arrive at Eq.
(\ref{1}), which is now to be understood as valid for each of the
three components of the displacement vector.
In particular, we note that the predictions of (I) for the
displacement and rotation angle distributions in the surroundings
of a flow event ($P(u)\propto u^{-5/2}$ and 
$P(\phi)\propto \phi^{-2}$) are valid also in the anisotropic
case.
The first prediction has recently been confirmed in computer
simulations of a binary Lennard-Jones mixture \cite{TBS99}, the
second is consistent with the small rotation angle distribution
tentatively inferred from NMR experiments by B{\"o}hmer and
Hinze on glycerol, $P(\phi)\propto 1/\sin^2(\phi)$ \cite{boh98}.

We now show that it is possible to characterize flow events
according to their anisotropy.
The elastic equilibrium in the surroundings of a flow event
region before as well as after the flow event implies that the
stress tensor change, $\Delta\sigma_{ij}$, has zero
divergence (i.e., obeys Eq.\ (\ref{2})).
Since ${\bf u}$ and $\Delta\sigma_{ij}$ are linked by linear
elasticity theory one has \cite{l+l} 

\begin{equation}\label{4}
\nabla ^2 \ ({\bf\nabla\cdot u})\ = 0\,.
\end{equation}
This equation can be solved asymptotically for
$r\rightarrow\infty$: 
Equation (\ref{1}) implies ${\bf\nabla\cdot u}\propto r^{-3}$.
Any real solution to the Laplace equation decaying as $r^{-3}$
can be written \cite{arf} as $\alpha\ P_2(\theta,\phi)/r^3$,
where $\alpha\geq 0$ is a constant and $P_2$ is a normalized
linear combination of second order spherical harmonics:
$P_2=\sum_{m=-2}^{m=2} c_m Y_{lm}$, where 
$c_m^*=c_{-m}$ and $\sum_{m=-2}^{m=2}|c_m|^2=1$.
This expression applies far away from the flow event: $r_0\ll r$,
where $r_0$ is the region size.
On the other hand, the expression does not apply beyond the
``solidity length'' $l$ discussed in (I), where essentially no
flow event induced displacements are expected.
Since ${\bf\nabla\cdot u}$ is dimensionless $\alpha$ has
dimension ${\rm (length)^3}$. 
Writing $\alpha=a/\rho_0$, where $\rho_0$ is the average (number)
density and $a$ is dimensionless, we have

\begin{equation}\label{5}
{\bf\nabla\cdot u} =\
a\ \frac{P_2(\theta,\phi)}{\rho_0\ r^3}\,(r_0\ll r\ll l)\,.
\end{equation}
The parameter $a$ is a measure of the flow event anisotropy,
the case $a=0$ corresponding to isotropic flow events.

In a homogeneous system described by linear elasticity theory the
density change following an elastic displacement is equal to
$-\rho_0 {\bf\nabla\cdot u}$ \cite{l+l}.
Thus, if a viscous liquid were homogeneous the density change in
the surroundings of a flow event would be given by Eq.\ (\ref{5})
(looking like an electronic d-orbital).  
However, the density of a viscous liquid is not quite spatially
constant.
As is easy to show, the density change induced by a flow event
has an extra term, $-{\bf u\cdot\nabla \rho}$, coming from the
fact that the whole density profile is displaced.
Far away from the flow event this extra density change term
dominates over the (${\bf\nabla\cdot u}$)-term.

The flow event induced changes given by Eqs. (\ref{1}) and
(\ref{5}) were calculated from the fact that there is a linear
relation between displacement and stress tensor change.
Therefore, these results are valid independent of the chemical
nature of the liquid.
One possible objection to these results is that dynamic
inhomogeneities most likely give rise to spatially varying
elastic constants.
However, being mainly interested in the high viscosity limit
where the solidity length is large, these inhomogeneities are not
expected to have any significant effect on the average
displacements in the surroundings of a flow event (the ``long
wave length'' limit).
Finally, we note that the sharp distinction between ``real''
motion and vibration is somewhat blurred by the fact that
``real'' motion takes place not only in the region itself in the
form of large jumps but also in the surroundings in the form of
small jumps.
However, as may be shown from Eq. (\ref{1}), the dominant
contribution to the mean-square displacement of a molecule comes
from the ``real'' motion of molecules inside regions.

To summarize, arguing from the ``solidity'' of viscous liquids,
the flow induced displacements in the surroundings have been
calculated for the general, anisotropic case.
It has been shown that the $r$-dependence of these displacements
is the same as that induced by isotropic flow events.
A dimensionless number $a$ has been introduced as a measure of
the degree of anisotropy of a flow event.

\acknowledgements
The author wishes to thank Austen Angell and Ralph Chamberlin for
numerous stimulating discussions and also for their most kind
hospitality during the author's stay at Arizona State University,
where parts of this work was carried out.
This work was supported by the Danish Natural Science Research
Council.


\begin{references}
\bibitem{note} In general, if a force field $F$ is applied to a
solid, the displacement is given by a standard Green's function
expression, 
$u_i({\bf r})=\int d{\bf r}' G_{ij}({\bf r}-{\bf r}')F_j({\bf
r}')$,
where $G_{ij}\propto r^{-1}$ (the precise $G_{ij}$ is given in
the above reference). This only applies for tranlationally
invariant solids and not for our case where the solid has a hole
in it, but far away from the hole our case is also
translationally invariant.
\end{references}
\end{document}